\documentclass[aps,prd,nofootinbib]{revtex4}
\usepackage{graphicx}

\begin{document}

\title{Upper limit on NUT charge from the observed terrestrial Sagnac effect}
\author{A. Kulbakova}
\email{kulbakova.a@mail.ru}
\affiliation{Zel'dovich International Center for Astrophysics, Bashkir State Pedagogical University, 3A, October Revolution Street, Ufa 450008, RB, Russia}
\author{R.Kh. Karimov}
\email{karimov\_ramis\_92@mail.ru}
\affiliation{Zel'dovich International Center for Astrophysics, Bashkir State Pedagogical University, 3A, October Revolution Street, Ufa 450008, RB, Russia}
\author{R.N. Izmailov}
\email{izmailov.ramil@gmail.com}
\affiliation{Zel'dovich International Center for Astrophysics, Bashkir State Pedagogical University, 3A, October Revolution Street, Ufa 450008, RB, Russia}
\author{K.K. Nandi}
\email{kamalnandi1952@rediffmail.com}
\affiliation{Zel'dovich International Center for Astrophysics, Bashkir State Pedagogical University, 3A, October Revolution Street, Ufa 450008, RB, Russia}
\affiliation{High Energy and Cosmic Ray Research Center, University of North Bengal, Siliguri 734 013, WB, India}

\begin{abstract}
The \textit{exact} Sagnac delay in the Kerr-Taub-NUT (Newman-Unti-Tamburino) spacetime is derived in the equatorial plane for non-geodesic as well as geodesic circular orbits. The resulting formula, being exact, can be directly applied to motion in the vicinity of any spinning object including black holes but here we are considering only the terrestrial case since observational data are available. The formula reveals that, in the limit of spin $a\rightarrow 0$, the delay does not vanish. This fact is similar to the non-vanishing of Lense-Thirring precession under $a\rightarrow 0$ even though the two effects originate from different premises. Assuming\ a reasonable input that the Kerr-Taub-NUT\ corrections are subsumed in the average residual uncertainty in the measured Sagnac delay, we compute upper limits on the NUT charge $n$. It is found that the upper limits on $n$ are far larger than the Earth's gravitational mass, which has not been detected in observations, implying that the Sagnac effect cannot constrain $n$ to smaller values near zero. We find a curious difference between the delays for non-geodesic and geodesic clock orbits and point out its implication for the well known "twin paradox" of special relativity.
\end{abstract}

\maketitle


\section{Introduction}

The Kerr-Taub-NUT spacetime is an axisymmetric, stationary solution of Einstein's vacuum field equations with mass ($M$), spin parameter ($a$) and NUT charge ($n$). The NUT charge is an additional feature of general relativity, which has been interpreted as the dual mass or gravitational analogue of a magnetic monopole in electrodynamics (see, e.g., \cite{1}). The NUT charge will be later expressed in relativistic units as its gravitational radius $\left( Gn\right) /c^{2}$ for comparing it with the gravitational radius of the Earth's mass $\left( GM_{\oplus }\right) /c^{2}$ under consideration. A lot of work on different implications of the NUT\ charge can be found in the literature (see, e.g., \cite{2,3,4,5,6,7,8,9,10}, but the list is by no means exhaustive).

Lynden-Bell and Nouri-Zonoz \cite{2} reviewed the dynamics of particles interacting with monopoles and were the first to initiate investigation of the observational possibilities for NUT charge. Kagramanova et al. \cite{3} calculated the phase shift for a charged particle interference experiment in a more general eletrovac Pleba\'{n}ski-Demia\'{n}ski black hole spacetime. In the same spacetime, in the linear approximation, Hackmann and L\"{a}mmerzahl \cite{4} derived an upper bound for the NUT parameter $n$ using the osculating orbital elements of Mercury. See also \cite{5}. Chakraborty and Majumdar \cite{1} investigated influence of NUT charge on Lense-Thirring (LT) precession frequency and showed that it remains non-zero even in the limit of $a\rightarrow 0$, a fact they attributed it to the existence of a "Copernican frame" (a.k.a. the fixed star Newtonian frame). The works in \cite{1,2,3,4} provide the motivation for the present paper seeking to place bounds on $n$ from a different experiment.

We shall use the experimental data from "around-the-world" type experiments. The early 1971 experiment by Hafele and Keating \cite{11} involved flying of two atomic clocks on board a commercial airliner in the east and westward directions circumnavigating the Earth. A more precise, and novel, 1985 experiment by Allan, Weiss and Ashby \cite{12} used electromagnetic signals transmitted from four GPS satellites instead of portable clocks observing a $90$ day run around the Earth. Schlegel \cite{13} has shown that the time asynchrony between two flying clocks is exactly the same as the time asynchrony between circumnavigating electromagnetic signals and this asynchrony is exactly the same as the Sagnac delay, to leading order.

The purpose of the present paper is to derive exact formulas for the Sagnac delay in the Kerr-Taub-NUT (KTN) spacetime and to estimate upper limits on the NUT charge $n$ by assuming an input that the KTN corrections to the flat space delay are less than the observed average residual error or uncertainty in the measured value of the delay. We shall show that a Copernican frame effect, similar to the one argued by Chakraborty and Majumdar \cite{1}, also appears in the form of a non-vanishing delay in the limit $a\rightarrow 0$. We shall point out that the difference in the delay for two types of orbits, non-geodesic and geodesic, in the flat space limit has an implication for twin paradox. Its implications for Mach's principle have been recently discussed elsewhere \cite{14,15}.

As a method of fixing the upper limit on $n$, we shall assume an input similar in spirit to the one adopted by Hackmann and L\"{a}mmerzahl \cite{4}. They used a bound on the conicity derived from the error tolerance of the inclination of Mercury's orbit around the Sun, which produced, in non-dimensionalized units, the limit $\left\vert n\right\vert \leq 0.032$. We shall assume that the the Kerr-Taub-NUT\ correction to the flat space Sagnac delay is subsumed in the average error residual reported by Allan, Weiss and Ashby \cite{12}. We shall also rely on the known fact that a spinning black hole metric adequately describes Earth's gravity in the weak field valid near its surface, where the circular motions take place (LT precession in the Kerr metric is an example).

The Sagnac effect briefly is as follows. Consider a circular turntable of radius $R$ having a light source/receiver (meaning the source \textit{and} the receiver at the same point) fixed to the turntable. A beam of light split into two at the source/receiver are made to follow the same closed path near the rim in opposite directions before they are re-united at the source/receiver. In the limit, if the turntable is not rotating, the beams will arrive at the same time at the source/receiver and an interference fringe will appear. When the turntable rotates with angular velocity $\omega_{0}$, the arrival times at the source/receiver will be different for co-rotating and counter-rotating beams: longer in the former case and
shorter in the latter. This difference in arrival times is called the Sagnac delay, which is measured by superimposing the two arriving beams with phase differences causing a shift in the interference fringes. The delay or fringe shift is a consequence of the lack of simultaneity (asynchrony)\ for motion of light signals along a closed loop.

The\ total arrival time lag between the two light beams, as measured at the source/receiver, can be obtained from special relativity, which gives, to first order in $\omega _{0}$,
\begin{equation}
\delta \tau _{S}=\frac{4\mathbf{\omega }_{0}\mathbf{.S}}{c^{2}},
\end{equation}%
where $\mathbf{S}$ is the area of the projection, orthogonal to the rotation axis,\ of the closed path followed by the waves contouring the turntable, $c$ is the speed of light in vacuum and $\mathbf{\omega }_{0}$ is the angular velocity of the turntable. Universal nature of this effect with different derivations exist, see, e.g., \cite{16,17,18}. The corresponding phase shift, $\delta \phi =\left( \frac{2\pi c}{\lambda }\right) \delta \tau _{S}$, has been accurately tested for various types of matter waves, see e.g., \cite{19}. Sagnac effect (1) is a special relativistic effect \cite{20}, with an extension to the quantum regime as discussed earlier by Anandan \cite{21}. One remarkable modern use of the effect lies in the global navigational systems, such as GPS, GLONASS etc, in which the rotation of Earth needs to be taken into account while using radio signals to synchronize clocks. This advantage was also used by Allan, Weiss and Ashby \cite{12} to arrive at more precise delay measurements.

We shall calculate general relativistic corrections to the Sagnac delay (1) due to mass, spin and NUT charge, when the "turntable" is a massive spinning compact object, the Earth and the source/receiver is a geostationary satellite. The effect has been previously worked out in different solutions of gravity, e.g., in the Kerr-Sen string metric \cite{22}, in the Brans-Dicke theory \cite{23,24}. However, the earliest work on Sagnac delay in general relativity, to our knowledge, was done by Ashtekar and Magnon \cite{25}. Tartaglia \cite{26} calculated the corrections to the delay due to mass and spin in the Kerr metric, and here we shall follow his methodology.

The paper is organized as follows: In Sec.2, we state the Kerr-Taub-NUT solution for a massive rotating compact object. In Section 3, we consider the equatorial circular motion of the source/receiver and in Sec.4, we estimate upper limits on the NUT charge $n$. In Sec.5, we consider the geodesic motion of GNSS clocks and its update to constrain $n$. Sec.6 concludes the paper. In the Appendix, we calculate the Sagnac delay for geodesic motion. We shall choose units such that $G=c=1$ unless specifically restored.

\section{Kerr-Taub-NUT metric}

The metric of the Kerr-Taub-NUT spacetime in Boyer-Lindquist coordinates $(x^{0}=t,x^{1}=r,x^{2}=\theta ,x^{3}=\phi )$ is
\begin{eqnarray}
ds^{2} &=& \frac{1}{\Sigma }(\Delta -a^{2}\sin ^{2}{\theta })dt^{2}-
\frac{2}{\Sigma }\left[\Delta A-a(\Sigma +aA)\sin ^{2}{\theta }\right] dtd\phi \nonumber\\
&&-\frac{1}{\Sigma }\left[ (\Sigma +aA)^{2}\sin ^{2}{\theta }-A^{2}\Delta %
\right] d\phi ^{2}-\frac{\Sigma }{\Delta }dr^{2}-\Sigma d\theta ^{2}.
\end{eqnarray}%
Here $\Sigma $, $\Delta $ and $A$ are defined by
\begin{eqnarray}
\Sigma &=& r^{2}+(n+a\cos ^{2}{\theta })^{2},\;\Delta =r^{2}-2Mr-n^{2}+a^{2}, \\
A &=& a\sin ^{2}{\theta }-2n\cos {\theta}.
\end{eqnarray}%
The parameters $(M,a,n)$ all have the same dimension of length in relativistic units. The source of the gravitational field has mass $M$, total angular momentum $J=Ma$ along the $z$ direction, and NUT charge $n$. The two solutions $r_{\pm }=M\pm \sqrt{M^{2}-a^{2}+n^{2}}$ of the equation $\Delta =0$ define the radii of the inner ($r_{-}$) and outer ($r_{+}$) horizons, when $a^{2}<M^{2}+n^{2}$. Our attention will be confined to the region outside the outer horizon: $r\geq r_{+}$.

\section{Equatorial orbit of source/receiver}

Following Tartaglia [26], consider that the source/receiver (geostationary satellite) is sending two oppositely directed light beams along a circumference on the equatorial plane $\theta =\pi /2$ of the rotating KTN black hole described by metric (2). Suitably placed mirrors send back to their origin both beams after a circular trip about the rotating central mass. Assume further that satellite is orbiting the central mass at a radius $r=R=$ const. far away from the horizon. Then the metric (2) reduces to
\begin{equation}
d\tau^{2} = \frac{R^{2}(1-2M/R)-n^{2}}{R^{2}+n^{2}}dt^{2}+\frac{4a(MR+n^{2})}{R^{2}+n^{2}}dtd\phi -\frac{(R^{2}+n^{2})^{2}+a^{2}(R^{2}+2MR+3n^{2})}{R^{2}+n^{2}}d\phi^{2}.
\end{equation}%
Assuming uniform axial rotation speed $\omega _{0}$ of the KTN black hole, the rotation angle $\phi _{0}$ of the satellite is given by
\begin{equation}
\phi _{0}=\omega _{0}t.
\end{equation}%
This yields
\begin{equation}
d\tau ^{2} = \frac{R^{2}(1-a^{2}\omega _{0}^{2}-R^{2}\omega_{0}^{2})-2MR(1-a\omega _{0})^{2} - n^{2}(1-4a\omega _{0}+3a^{2}\omega _{0}^{2}+2R^{2}\omega_{0}^{2}) - n^{4}\omega _{0}^{2}}{R^{2}+n^{2}}dt^{2}.
\end{equation}%
For light moving along the same circular paths it must obey $d\tau =0$. Assuming $\Omega $ to be the angular velocity of light motion along the paths, we have
\begin{equation}
-R^{2}+2MR+n^{2}-4a(MR+n^{2})\Omega +\left[(R^{2}+n^{2})^{2}+a^{2}(R^{2}+2MR+3n^{2})\right] \Omega ^{2}=0.
\end{equation}%
Solving the quadratic Eq.(8), one finds two roots that represent the angular velocity $\Omega_{\pm}$ of light for the co- and counter rotating motion, given by
\begin{equation}
\Omega _{\pm }=\frac{2a(MR+n^{2})\pm \sqrt{(R^{2}+n^{2})^{2}(R^{2}-2MR+a^{2}-n^{2})}}{(R^{2}+n^{2})^{2}+a^{2}(R^{2}+2MR+3n^{2})}.
\end{equation}%
The rotation angles $\phi _{\pm }$ for light then are
\begin{equation}
\phi _{\pm }=\Omega _{\pm }t.
\end{equation}%
Eliminating $t$ between Eqs.(6) and (10), we obtain
\begin{equation}
\phi _{\pm }=\frac{\Omega _{\pm }}{\omega _{0}}\phi _{0}.
\end{equation}

The first intersection of the world lines of the two light rays with the one of the orbiting source/receiver after the emission at time $t=0$ is, when the angles are
\begin{eqnarray}
\phi_{+}=\phi _{0}+2\pi, \\
\phi_{-}=\phi _{0}-2\pi,
\end{eqnarray}%
which give
\begin{equation}
\frac{\Omega_{\pm }}{\omega _{0}}\phi _{0}=\phi _{0}\pm 2\pi .
\end{equation}%
Solving for $\phi _{0}$,
\begin{equation}
\phi _{0\pm }=\mp \frac{2\pi \omega _{0}}{\Omega _{\pm }-\omega _{0}},
\end{equation}%
we have, putting the expressions from (9),
\begin{equation}
\phi _{0\pm }=\mp \left( 2\pi \omega _{0}\right) /\left[ \frac{%
2a(MR+n^{2})\pm \sqrt{(R^{2}+n^{2})^{2}(R^{2}-2MR+a^{2}-n^{2})}}{%
(R^{2}+n^{2})^{2}+a^{2}(R^{2}+2MR+3n^{2})}-\omega _{0}\right] .
\end{equation}

The proper time of the rotating source/receiver, deduced from Eq.(7) using Eq.(6), is
\begin{equation}
d\tau =\left[\frac{R^{2}-2MR-n^{2}+4a(MR+n^{2})\omega _{0} - \left\{(R^{2}+n^{2})^{2}+a^{2}(R^{2}+2MR+3n^{2})\right\}
\omega_{0}^{2}}{R^{2}+n^{2}}\right]^{\frac{1}{2}}\frac{d\phi_{0}}{\omega_{0}}.
\end{equation}%
Finally, integrating between $\phi _{0-}$ and $\phi _{0+}$ , we obtain the exact Sagnac delay
\begin{equation}
d\tau =\left[\frac{R^{2}-2MR-n^{2}+4a(MR+n^{2})\omega _{0} - \left\{(R^{2}+n^{2})^{2}+a^{2}(R^{2}+2MR+3n^{2})\right\}
\omega_{0}^{2}}{R^{2}+n^{2}}\right]^{\frac{1}{2}}\frac{\phi _{0+}-\phi_{0-}}{\omega _{0}}.
\end{equation}
Using the integration limits from Eq.(16), we explicitly write the exact
formula as
\begin{eqnarray}
\delta \tau _{S\;\textmd{\scriptsize{nongeo}}}^{\textmd{\scriptsize{KTN}}} &=&-\left(\frac{4\pi}{R}\right)
\left[2a(MR+n^{2})-(R^{2}+n^{2})^{2}\omega_{0}-a^{2}(R^{2}+2MR+3n^{2})\omega_{0}\right] / \nonumber\\
&&\left[ (1+n^{2}/R^{2})\left\{ R^{2}-2MR-n^{2}+4a(MR+n^{2})\omega _{0}\right.\right. \nonumber\\
&&\left. \left. -\{(R^{2}+n^{2})^{2}+a^{2}R(R+2M)+3n^{2}a^{2}\}\omega_{0}^{2}\right\} ^{1/2}\right] \\
&=&\delta \tau _{S}+ \textmd{correction terms},
\end{eqnarray}
Eq.(19) is the exact master formula for the Sagnac delay for nongeodesic source/receiver motion that we had promised. Its leading order term obtained by setting $M=0$, $a=0$, $n=0$ is just the flat space term $\delta \tau_{S}=4\pi R^{2}\omega _{0}$ as in (1), which is often interpreted as the gravitational analogue of the Bohm-Aharonov effect \cite{27} although the light beams are not truly moving in the gravitation free space. The best situation that possibly comes closer to the Bohm-Aharonov effect could be developed with light beams moving along a flat space torus (see for details, Semon \cite{28}). Nevertheless, as shown by Ruggiero \cite{29}, Eq.(1) completely agrees with the one of the gravito-electromagnetic Bohm-Aharonov interpretation \cite{30}. For the viewpoint of Bohm-Aharonov quantum interference in general relativity, see \cite{31,32}.

A Post-Newtonian first order approximation for a static observer sending a pair of light beams in opposite directions along a closed triangular circuit, instead of a circle, was worked out by Cohen and Mashhoon \cite{33} and they found in that approximation the same result\ as (1). So what is important is not the shape but the closedness of the orbit. This important information will allow us to consider an \textit{equivalent} circular orbit in the sequel, no matter what the shape of the enclosed area is.

The Taub-NUT spacetime obtained in the limit $a\rightarrow 0$ has been investigated quite well. For instance, geodesic structures in that spacetime have been analyzed in detail in \cite{34}. From (19), in the same limit, one finds that the Sagnac delay is not vanishing:
\begin{equation}
\delta \tau _{\textmd{\scriptsize{nongeo}}}^{\textmd{\scriptsize{TN}}} = \delta \tau _{S\;\textmd{\scriptsize{nongeo}}}^{\textmd{\scriptsize{KTN}}} |_{a=0}=\frac{4\pi (R^{2}+n^{2})^{\frac{3}{2}}\omega _{0}}{\sqrt{%
R^{2}-2MR-n^{2}-(R^{2}+n^{2})^{2}\omega _{0}^{2}}}.
\end{equation}%
This is the delay measured by a source/receiver moving with an angular velocity $\omega _{0}$ around a static source with a NUT charge $n$. As explained by Chakraborty and Majumdar \cite{1}, the Taub-NUT spacetime is not invariant under time reversal $t\rightarrow -t$, indicating some sort of "rotation" analogous to a electrodynamic magnetic monopole. They argue that it is this analogue rotation that is responsible for the nonvanishing of the LT precession. The interesting thing here is that the same arguments apply to the nonvanishing of Sagnac delay under $a\rightarrow 0$ as well, even though the effects have completely different physical origin - LT precession is caused by frame dragging, while Sagnac delay is caused by the time asynchrony.

The quadratic equation in $n^{2}$ under the radical sign in the denominator of Eq.(19) yields two exact roots
\begin{equation}
n_{\pm }^{2}=\frac{\left( 1-4a\omega _{0}+3a^{2}\omega _{0}^{2}+2R^{2}\omega
_{0}^{2}\right)}{2\omega _{0}^{2}} \pm \frac{\left( a\omega _{0}-1\right) \sqrt{1-6a\omega
_{0}+9a^{2}\omega _{0}^{2}-8MR\omega _{0}^{2}+8R^{2}\omega _{0}^{2}}}{2\omega _{0}^{2}}.
\end{equation}%
In order that the denominator in (19) be not imaginary (which we thus call "reality constraint"), one must ensure that $n^{2}<n_{\pm }^{2}$. We shall soon evaluate $n_{\pm }$ in the case of motion around the Earth. In the meantime, for $a=0,n=0,$ one ends up with the Schwarzschild black hole and from Eq.(19) the Sagnac delay then follows as
\begin{eqnarray}
\delta \tau |_{a=n=0} &=& \frac{4\pi R^{2}\omega _{0}}{\sqrt{%
1-2M/R-R^{2}\omega _{0}^{2}}} \\
&\simeq& 4\pi R^{2}\omega _{0}+4\pi MR\omega _{0}+2\pi R^{4}\omega _{0}^{3}.
\end{eqnarray}%
The second term represents the correction due purely to mass. At $M\rightarrow 0$, one readily recovers, to leading order, the flat space delay $\delta \tau _{S}=4\pi R^{2}\omega _{0}$, corrections to which will be obtained directly from the exact expression (19) in the next section. When $a=0$, $M=0$, but $n\neq 0$ in (19), we find that there is a correction to $\delta \tau _{S}$ purely due to $n$, that is\textit{\ independent} of radius $R$, given by the expression
\begin{equation}
\delta \tau |_{a=M=0}=4\pi R^{2}\omega _{0}+6\pi n^{2}\omega _{0}.
\end{equation}%
However, the term still depends on the "radius" of the NUT charge $n$, which is equal to the reduced event horizon $r_{\pm }=\pm n$ and, of course, on the orbital angular speed $\omega _{0}$ of the source/observer.

\section{Upper limits from the terrestrial Sagnac data}

We shall consider the result of the precision experiment by Allan, Weiss and Ashby \cite{12} measuring the terrestrial Sagnac delay. Instead of portable clocks, they used four GPS satellites\ transmitting electromagnetic signals that can have a common view from remote stations on Earth. The experiment is equivalent to $90$ day independent runs yielding flat space one-way delays from about $240$ to $350$ ns with Sagnac error residual of only $5$ ns. (One way delay is $\frac{1}{2}\delta \tau _{S}$, which means that there is a stationary clock on Earth, and its reading is compared with that of the airborne clock after its non-geodesic circumnavigation). The projected area on Earth in the experiment is not circular but, following Mashhoon and Cohen \cite{33}, we can always calculate an equivalent radius yielding the same measurements. Any of the measured values of $\frac{1}{2}\delta \tau _{S}$ can be used in the expression (1) to obtain the desired equivalent radius. Allan, Weiss and Ashby \cite{12} used GPS satellites around Earth moving with Earth's angular speed $\omega _{0}$ given by (27).

The equivalent radius $R_{\textmd{\scriptsize{eq}}}$ (yielding one of the measured values, say, $\frac{1}{2}\delta \tau _{S}=240$ ns, assuming a conventional circumnavigation) and the other Earth data are as follows:
\begin{equation}
R\rightarrow R_{\textmd{\scriptsize{eq}}}=7\times 10^{6}\;\textmd{m},
\end{equation}%
\begin{equation}
\omega _{0}=\Omega _{\oplus }=7.30\times 10^{-5} \;\textmd{rad/s} \Rightarrow
2\omega _{0}/c^{2}=1.6222\times 10^{-21}\;\textmd{rad} \left(\textmd{s/m}^{2}\right),
\end{equation}%
\begin{equation}
M\rightarrow GM_{\oplus }/c^{2}=4.40\times 10^{-3}\;\textmd{m},
\end{equation}%
\begin{equation}
a=a_{\oplus }=9.81\times 10^{6}\;\textmd{m$^{2}$/s},
\end{equation}%
\begin{equation}
c=3\times 10^{8}\;\textmd{m/s}.
\end{equation}%
The flat space zeroth order Sagnac delay $\delta\tau_{S} = 4\pi\omega_{0}R^{2}/c^{2}$, with $\omega_{0}=\Omega_{\oplus}$, $R=R_{\oplus}=6.371\times 10^{6}$ m, due to the east and westward equatorial motion of the airborne atomic clocks, works out to
\begin{equation}
\delta \tau _{S}=2\times \frac{2\Omega _{\oplus }}{c^{2}}\times \pi
R_{\oplus }^{2}=4.148\times 10^{-7}\;\textmd{s} =2\times 207.4\;\textmd{ns}.
\end{equation}%
As well known, this famous value $\frac{1}{2}\delta \tau _{S}$ ($=207.4$ ns) is the one way delay measured by Hafele and Keating with error residual $\sim 10$ nsec in their experiment \cite{33}.

We plug the above Earth data (26)-(30) into $\frac{1}{2}\delta\tau$ of exact Eq.(19), restoring $\omega_{0}\rightarrow \omega_{0}/c^{2}$, $a\rightarrow a_{\oplus }/c^{2}$, and converting second to nanosecond (ns) scale, $\frac{1}{2}\delta \tau \rightarrow \frac{1}{2}\delta \tau \times 10^{9}$ ns. Then expanding, we find%
\begin{eqnarray}
\frac{1}{2}\delta \tau \times 10^{9} &=&\left( 239.459+7.30\times 10^{-12}n^{2}+8.75\times 10^{-26}n^{4}+...\right) \;\textmd{ns} \\
&=&\frac{1}{2}\delta\tau_{S}^{\textmd{\scriptsize{flat}}} + \textmd{ corrections},
\end{eqnarray}%
where $n$ has the dimension of length but here it is understood to have been non-dimensionalized as $n$ m$^{-1}$. This expansion here is meant only to show that the first term is the experimentally observed flat space value $\frac{1}{2}\delta\tau_{S}^{\textmd{\scriptsize{flat}}}$ ($\equiv 239.718$ ns) and the remaining terms are the corrections due to $n$. If we subtract $\frac{1}{2}\delta \tau _{S}$ from the total $\frac{1}{2}\delta \tau $ of (19), then what remains would be just the total correction term due to $M,a$ and $n$.

Since the Earth values of $M$ and $a$ are already plugged in Eq.(19) before expansion, only $n$ is appearing in (32). Using the assumption that the total correction term is less than or equal to the average error residual of $5$ ns, we obtain%
\begin{equation}
n\leq 8.24\times 10^{5} \textmd{ m}.
\end{equation}%
The same limit can be obtained also by using\ the other end value $\frac{1}{2}\delta\tau_{S}^{\textmd{\scriptsize{flat}}}$ ($\equiv 350$ ns). On the other hand, the reality constraint $n^{2}<n_{\pm }^{2}$ yields
\begin{equation}
n_{+}^{2}=-1.52\times 10^{42} \textmd{ m}^{2},\quad n_{-}^{2}=4.9\times 10^{13} \textmd{ m}^{2},
\end{equation}%
whence $n_{+}$ is ruled out for being imaginary but
\begin{equation}
n<n_{-}=7\times 10^{6} \textmd{ m}.
\end{equation}%
The limits in (34) and (36) on the gravitational radius of the NUT charge $n$ far exceed the gravitational mass of the Earth ($=4.40\times 10^{-3}$ m), and has never been observed in any terrestrial experiment. In principle, however, the limits do allow $n\sim 0$, but there is no known physical criterion by which it can be achieved. So, there is no hope to constrain $n$ by terrestrial Sagnac delay experiment to a value near zero even if the level of accuracy is raised from nanosecond to unlikely femtosecond level.

\section{Global Navigation Satellite Systems (GNSS) update}

It has been brought to our notice\footnote{We are indebted to an anonymous referee for pointing it out.} that updated data on clock synchronization using two-way radio links between two ground stations and a freely falling (geodesic motion) satellite in the GNSS are available. The satellite systems are equipped with accurate, stable atomic clocks on-board, while there are precision clocks fixed on the ground providing world-wide access to position, velocity and time of all events. An excellent recent review by Ashby \cite{35} enumerates the various relativistic factors that have to be accounted for if the systems have to work well. These factors include relativistic principles, concepts and effects such as the constancy of the speed of light, relativity of synchronization, coordinate time, proper time, time dilation, the Sagnac effect, the weak equivalence principle and gravitational frequency shifts. Additionally, Shapiro time delay and tidal effects caused by the moon and the sun might also be corrected for in the future experiments. See \cite{36} for more details.

We shall use not the basic Sagnac delay itself but only the updated fluctuation in the delay as observed in the GNSS clock system after circumnavigation. The fluctuation is caused by the jittery motion of the satellite about the equator. When the positions of the Earth stations are fixed and the satellite moves strictly over the equator, the basic Sagnac delay is constant. In reality, however, the actual position of the geostationary satellite varies slightly over a $24-$hr period. A study demonstrated that the time varying fluctuation in the delay due to satellite jitter could add more than $0.5$ ns to the constant basic value \cite{37}.

A crucial point about the relevant radius is to be noted here. In the previous section, the identification $\omega _{0}=\Omega _{\oplus}=7.27\times 10^{-5}$ rad/s in (27) is incomplete in the sense that it is not related to a force-balance equation
\begin{equation}
\Omega _{\oplus }=\sqrt{GM_{\oplus }/R_{\textmd{\scriptsize{geost}}}^{3}},
\end{equation}%
that yields a unique geostationary radius $R_{\textmd{\scriptsize{geost}}}=6.62R_{\oplus}=4.2176\times 10^{6}$ m. This incompleteness left the radius of the orbiting clock to be freely chosen including the choice of non-geodesic orbits ($R=R_{\oplus }$) as occurred in the Hafele-Keating (HK) experiment, where clocks are transported round the Earth by engine-driven aircrafts \cite{11} or $R=R_{\textmd{\scriptsize{eq}}}$ \cite{12}. On the other hand, GNSS\ clocks necessarily follow a geodesic orbit occurring exactly at the equatorial geostationary radius $R=R_{\textmd{\scriptsize{geost}}}$.

The non-trivial changes in the two-way Sagnac delay due to the change \ from non-geodesic to geodesic orbit can in fact be quite large:%
\begin{equation}
\delta \tau _{S\;\textmd{\scriptsize{nongeo}}}^{\textmd{\scriptsize{HK}}}=\frac{4\pi \Omega_{\oplus }R^{2}}{c^{2}}=414.8 \;\textmd{ns},\;\left(\omega _{0}=\Omega _{\oplus },\; R=R_{\oplus}\right) ,
\end{equation}
\begin{equation}
\delta \tau _{S\;\textmd{\scriptsize{geo}}}=\frac{4\pi \Omega _{\oplus }R^{2}}{c^{2}}=18044.6\;\textmd{ns},\;\left( \omega _{0}=\Omega _{\oplus },\; R=R_{\textmd{\scriptsize{geost}}}\right) .
\end{equation}%
This shows that the delay from the geodesic motion (39) would be $43.7$ times larger than that from the non-geodesic motion (38). It can be seen that by putting $\Omega _{\oplus }$ from Eq.(37) into Eq.(39), one obtains
\begin{equation}
\delta \tau _{S}^{\textmd{\scriptsize{geo}}}=4\pi \sqrt{M_{\oplus }R_{\textmd{\scriptsize{geost}}}} ,
\end{equation}%
which is just the redefined Newtonian expression for the delay. This term will appear in the leading order in the weak field expansion of exact general relativistic master expression for the Sagnac delay (A.9) for geodesic motion in the KTN metric. This is derived in the Appendix. Letting the parameters assume the Earth values $a=a_{\oplus }$, $M=M_{\oplus }$ and the geodesic orbit $R=R_{\textmd{\scriptsize{geost}}}$ in Eq.(A.9), and converting second to nanosecond, we find
\begin{eqnarray}
\left\vert \delta \tau _{S\pm\:\textmd{\scriptsize{geo}}}^{\textmd{\scriptsize{KTN}}}\right\vert
&=&\delta \tau _{S\pm\:\textmd{\scriptsize{geo}}}^{\textmd{\scriptsize{Kerr}}}+\;\textmd{corrections}\:O(n^{2})
\\
&=&[\left( \pm 1.80\times 10^{4}\mp 1.43\times 10^{-10}\right) +7.74\times 10^{-3}n^{2}+...] \;\textmd{ns},
\end{eqnarray}%
where, from Eq.(A.11), $\delta \tau _{S\pm\:\textmd{\scriptsize{geo}}}^{\textmd{\scriptsize{Kerr}}}=\left(\pm 1.80\times 10^{4}\mp 1.43\times 10^{-10}\right) $ ns, the first term being contributed by $M_{\oplus }$ and the second by $a_{\oplus }$. The remaining term $7.74\times 10^{-3}n^{2}$ in Eq.(42) expresses the correction caused by the unknown NUT charge $n$.

Like before, our idea is to identify the above correction with the fluctuation or uncertainty in the observed data on the round-trip Sagnac delay. In this context, we note that precision measurement of the Two-Way Satellite Time and Frequency Transfer (TWSTFT) highly depends on the residual non-reciprocity delays - one of them is caused by the Sagnac delay or synchronization discontinuity found between the re-united flying clocks \cite{13}. GNSS calibration of clocks by means of TWSTFT involving Sagnac delay has achieved an accuracy at the level of nanosecond \cite{37,38}.

Tseng \textit{et al. }\cite{39} showed that the Sagnac delay can be calculated using the Earth station coordinates and the actual ephemeris data. Their experiment involved Earth stations located at the National Institute of Information and Communications Technology (NICT) in Japan and the Telecommunication Laboratories (TL) in Taiwan. This experiment on the time rate of variation of the Sagnac delay, called diurnal \cite{39}, predicted\ a variation $\Delta (\delta\tau _{\textmd{\scriptsize{S}}})$ of magnitude $\pm 0.25$ ns, which provides an update over $\sim 5$ ns obtained in the 1984 experiment \cite{12}. This updated value from TWSTFT then provides an upper limit on the correction term in Eq.(42).Thus, we have%
\begin{equation}
n <\sqrt{\frac{\Delta (\delta\tau _{\textmd{\scriptsize{S}}})}{7.74\times 10^{-3}}}\;\textmd{m} = 5.68\;\textmd{m},
\end{equation}%
which is far less than the one obtained in (36) but still a thousand times larger than the gravitational radius of the Earth's mass, Eq.(28). Such a huge amount of NUT charge $n$ of the Earth has not been detected by observations and so its existence is practically ruled out.

In the flat space limit from Eq.(19), we obtained $\delta \tau _{S\;\textmd{\scriptsize{nongeo}}}^{\textmd{\scriptsize{KTN}}}\left\vert _{M=0,a=0,n=0}\right. =\frac{4\pi R^{2}\omega _{0}}{c^{2}}\simeq 240$ ns, whereas from (A.8), we obtain $\delta \tau _{S\pm\:\textmd{\scriptsize{geo}}}^{\textmd{\scriptsize{KTN}}}\left\vert _{M=0,a=0,n=0}\right. =0$. Hence, we argue that the non-zero delay $\frac{4\pi R^{2}\omega _{0}}{c^{2}}$ or synchrony gap is caused by terrestrial inertial forces acting on the flying satellite clocks due to their non-geodesic motion, since such forces are not balanced by Earth's gravity. This observation has implications for the well known "twin paradox", which is no paradox but a genuine prediction of special relativity. The synchrony gap between the flying clocks brought about by the inertial forces is exactly the age difference between the circumnavigating twins (neglecting velocity effects) \cite{13,14,15}. The synchrony gap brought about by the inertial forces is exactly the age difference between the circumnavigating twins (neglecting velocity effects) \cite{13}. On the other hand, twins or clocks circumnavigating geodesically in flat space ($M=0,a=0,n=0$) will \textit{not} have any age difference between them upon reuniting.

\section{Conclusions}

In this paper, we derived the exact formulas (19) and (A.9) for Sagnac delay in the equatorial plane respectively for non-geodesic (arbitrary radius) and geodesic (geostationary radius) orbits around a  Kerr-Taub-NUT black hole. The delay is caused by an inevitable time asynchrony or time discontinuity exhibited by two on-board clocks upon reuniting after circumnavigation. There are three clear conclusions from the analyses, as summarized below:

(1) We notice that in the limit $a\rightarrow 0$, the Sagnac delay does not vanish similar to the non-vanishing of LT precession shown by Chakraborty and Majumdar \cite{1}, even though the two effects have totally different origin. This non-vanishing of the delay at $a\rightarrow 0$ shows that the source spin effect has not disappeared altogether since $n$ can be interpreted as having some kind of "rotation" of the NUT charge indicated by the non-invariance of Taub-NUT metric under time reversal $t\rightarrow -t$.

(2) As to the fixing of the upper limit on $n$ from experiment, the steps followed are as follows: First, we plugged the numerical Earth data directly into the exact formula (19), and expanded it to verify that the first term on the right side in Eq.(32) is indeed the observed flat space one-way value of the Sagnac delay $\frac{1}{2}\delta\tau_{S}$. Equating this value with the one reported in \cite{12}, we obtained an equivalent orbit radius $R_{\textmd{\scriptsize{eq}}}$, which is now used in the expanded form. Second, we computed the difference $\left( \frac{1}{2}\delta\tau -\frac{1}{2}\delta \tau_{S}\right)$, which contained only the corrections due to $M, a, n$ in closed form. Third, with the input assumption that $\left( \frac{1}{2}\delta \tau -\frac{1}{2}\delta\tau_{S}\right) \leq 5$ ns, which is the error residual observed in the Allan, Weiss and Ashby \cite{12} experiment, we obtained rather large upper limits (34,36) on the gravitational radius of the NUT charge $n$ that far exceed the gravitational mass of the Earth. Though the value $n\sim 0$ is in principle allowed by the limits, no independent physical argument to reach it is known to us. Even if the accuracy of observation is raised to femtosecond level (very unlikely), one cannot reduce the upper limit to smaller than $10^{2}$ m, still far larger than Earth's gravitational mass. Such huge NUT charge associated with Earth ought to have been otherwise detected, which has not been reported. Thus the conclusion is that the terrestrial delay experiment cannot constrain $n$ to acceptable values near zero unlike, e.g., obtained by Hackmann and L\"{a}mmerzahl \cite{4} using Mercury orbit data.

(3) We point out that both the Hafele-Keating \cite{11} and Allan-Weiss-Ashby \cite{12} experiments involve only non-geodesic orbits in the sense that the radius $R$ of the orbiting clocks does not obey the force-balance condition (i.e., Kepler's third law) for geodesic motion. The balance condition $\Omega_{\oplus}^{2}R = \frac{GM_{\oplus}}{R^{2}}$ ensuring geodesic orbit yields a geostationary orbit radius $R = R_{\textmd{\scriptsize{geost}}} = 6.62R_{\oplus} = 4.2176\times 10^{7}$ m, which is \textit{larger} than those used in \cite{11,12}. It was also shown in Eqs.(38,39) that the delays significantly differ depending on the type of the orbit. The force-balance condition has been shown to hold in the weak field limit of the KTN gravity, see (A.11). This condition has been used in the TWSTFT experiments in \cite{39}, where $R_{\textmd{\scriptsize{geost}}}=6.62R_{\oplus }$ was considered. If the satellite strictly follows the equatorial orbit, the Earth observer would see the satellite at a fixed place for 24-hrs and the round-trip Sagnac delay would be constant. But in practice, the satellite motion is jittery in its orbit causing fluctuations, called diurnal, that add small corrections to the ideal constant value. We used an updated range of fluctuation ($\pm 0.25$ ns) \cite{39} to constrain the NUT charge $n$. The upper limit is considerably improved as shown in (43) but unfortunately is still far larger than the Earth's mass. So the conclusion is that $n$ cannot be meaningfully constrained by the Sagnac delay observations.

Finally, as an interesting spin-off, it is argued that twins circumnavigating along non-geodesic paths will age differently, while those moving along geodesics will age similarly in the flat spacetime limit given by $\delta \tau _{S\;\textmd{\scriptsize{nongeo}}}^{\textmd{\scriptsize{KTN}}} \left\vert_{M=0,a=0,n=0}\right. =\frac{4\pi R^{2}\omega _{0}}{c^{2}}\simeq 240$ ns (where $R\neq R_{\textmd{\scriptsize{geost}}}$) and $\delta \tau _{S\pm\:\textmd{\scriptsize{geo}}}^{\textmd{\scriptsize{KTN}}}\left\vert _{M=0,a=0,n=0}\right. =0$ (independent of $R$). Even though the limit is mathematically correct, the argument of similar aging seems\ physically untenable. The reason is that, in the flat space limit, the geodesics are supposed to be only straightlines and circular paths should necessarily incur artificial forces. There cannot be any geodesic circular path in flat space. Hence the conclusion is that aging is caused only by the \textit{terrestrial inertial forces} acting on the clocks moving along non-geodesic circular paths.

\section*{Acknowledgement}

We are indebted to Arunava Bhadra for useful conversations. We thank anonymous referees for useful suggestions.
The reported study was funded by RFBR according to the research project No.18-32-00377.

\appendix
\section{Sagnac delay for geodesic equatorial orbit}

We consider a circular geodesic orbit of the source/reciver (freely falling satellites) at some arbitrary radius on the equator ($\theta =\pi /2$) sending light signals circumnavigating the Earth. Defining the velocity four-vector $\dot{x}^{\nu }=\frac{dx^{\nu }}{d\tau }$, the Lagrangian can be written as%
\begin{equation}
L=\frac{1}{2}g_{\mu \nu }\dot{x}^{\mu }\dot{x}^{\nu }
\end{equation}%
and the Euler-Lagrange $r-$equation is%
\begin{equation}
\frac{d}{d\tau }\left( \frac{\partial L}{\partial \dot{r}}\right) =%
\frac{\partial L}{\partial r}.
\end{equation}%
Since in metric (2), $g_{r\mu }=0$ for $r\neq \mu $, we have%
\begin{equation}
\frac{d}{d\tau }\left( g_{rr}\dot{r}\right) =\frac{1}{2}g_{\mu \nu ,r}%
\dot{x}^{\mu }\dot{x}^{\nu }.
\end{equation}%
Circular orbits are defined by the conditions%
\begin{equation}
\dot{r}=\ddot{r}=0,
\end{equation}%
and the Eq.(A.3) yields%
$$g_{tt,r}\dot{t}^{2}+2g_{t\phi ,r}\dot{t}\dot{\phi} + g_{\phi \phi ,r}\dot{\phi }^{2}=0.$$
Defining $\omega =\dot{\phi }/\dot{t}$, this equation yields the quadratic equation
\begin{equation}
g_{\phi \phi ,r}\omega ^{2}+2g_{t\phi ,r}\omega +g_{tt,r}=0.
\end{equation}%

From the metric (2), putting $dr=0$ at $r=R=$ const. and $d\theta =0$ at $\theta =\pi /2$, we find%
$$d\tau ^{2}=g_{tt}dt^{2}+2g_{t\phi }dtd\phi +g_{\phi \phi }d\phi^{2},$$
where
\begin{eqnarray}
g_{tt}&=&1-\frac{2(MR+n^{2})}{R^{2}+n^{2}},\quad g_{t\phi }=\frac{2a(MR+n^{2})}{R^{2}+n^{2}}, \nonumber\\
g_{\phi\phi}&=&-\frac{(R^{2}+n^{2})^{2}+a^{2}(R^{2}+2MR+3n^{2})}{R^{2}+n^{2}}.
\end{eqnarray}
The satellite's orbital speeds $\omega_{\pm}$ then follow from the two roots of quadratic Eq.(A.5). Using Eqs.(A.6), we obtain
\begin{equation}
\omega _{\pm } =\left[ \frac{an^{2}(M-2R)-aMR^{2}\pm P}{Q}\right],
\end{equation}
$$P \equiv \sqrt{MR^{7}+n^{2}R^{5}(M+2R)+n^{4}R^{3}(4R-M)+n^{6}R(2R-M)},$$
$$Q \equiv R^{5}-Ma^{2}R^{2}+n^{2}(2R^{3}+n^{2}R-2a^{2}R+Ma^{2}),$$
which shows that the angular speed $\omega _{\pm }$ of the satellite is determined by the metric itself, which now involves not only $M$ but also $a$ and $n$.

To fix the geodesic radius $R$, we customarily go over to the weak field static Newtonian limit, $a=0,n=0$ and identify $\omega_{\pm} = \Omega_{\oplus}$, $M=M_{\oplus }$. In that limit, from (A.7), we exactly recover Newtonian force-balance equation
\begin{equation}
\Omega _{\oplus }=\sqrt{GM_{\oplus }/R^{3}},
\end{equation}
that yields $R=R_{\textmd{\scriptsize{geost}}}=6.62R_{\oplus }$. However, this is no surprise as we are already considering general relativistic geodesic or force-free orbits. In the sequel, the notation $R$ is to be understood as $R_{\textmd{\scriptsize{geost}}}$. The exact Sagnac delay for geodesic motion then is $\delta \tau _{S\pm\:\textmd{\scriptsize{geo}}}^{\textmd{\scriptsize{KTN}}} = 4\pi \omega_{\pm}R^{2}$, which can be written out explicity as below:%
\begin{equation}
\delta \tau _{S\pm\:\textmd{\scriptsize{geo}}}^{\textmd{\scriptsize{KTN}}} =\pm 4\pi \left[ \frac{
-aMR^{2}\pm \sqrt{MR^{7}}}{R^{3}-a^{2}M}+\frac{n^{2}S}{2MR^{2}(R^{3}-a^{2}M)^{2}}+O(n^{4})\right],
\end{equation}
$$S \equiv 6aM^{2}R^{5}-4aMR^{6}+\sqrt{MR^{7}}\left( \mp 3a^{2}M^{2}\pm 2a^{2}MR\mp 3MR^{3}\pm 2R^{4}\right).$$
When the Kerr spin $a=0$, but $n\neq 0$ (Taub-NUT case), we obtain the Sagnac delay as%
\begin{equation}
\delta \tau _{S\pm\:\textmd{\scriptsize{geo}}}^{\textmd{\scriptsize{TN}}}=4\pi R^{2}\left[\pm \frac{\sqrt{MR^{7}+n^{2}R^{5}(M+2R)+n^{4}R^{3}(4R-M)+n^{6}R(2R-M)}}{R^{5}+2R^{3}n^{2}+Rn^{4}}\right].
\end{equation}%
The Kerr terms follow from the expansion of (A.9) at $n=0$ under the weak field conditions $\frac{a^{2}}{R^{2}}\ll 1$, $\frac{M}{R}\ll 1$, the resulting then coincides to leading order with the formula derived by Lichtenegger and Iorio \cite{14,26}:
\begin{equation}
\delta \tau _{\pm }^{\textmd{\scriptsize{Kerr}}}=\pm 4\pi \sqrt{MR}\mp 4\pi a\left( \frac{M}{R}\right) +...
\end{equation}%
and with $n=0$, we get the Schwarzschild value
\begin{equation}
\delta \tau _{S\pm }^{\textmd{\scriptsize{Sch}}}=\pm 4\pi \sqrt{MR},
\end{equation}%
which is Eq.(40) in the text.

\end{document}